\documentclass[fleqn]{article}
\usepackage[utf8]{inputenc}

\usepackage{amsmath}
\usepackage{mathrsfs}
\usepackage{rotating}
\usepackage{graphicx}
\usepackage{wrapfig}
\usepackage{float}
\usepackage[colorlinks]{hyperref}

\title%
{Maximization of the olfactory receptor neuron selectivity in the sub-threshold regime}%
\author{O.K.~Vidybida\thanks{%
Bogolyubov Institute for Theoretical Physics, Nat.
Acad. of Sci. of Ukraine,
14b, Metrolohichna Str., Kyiv 03143, Ukraine; url: http://www.vidybida.kiev.ua/}
}

\begin{document}
\maketitle

\begin{abstract}
It is known that if odors are presented to an olfactory receptor
neuron (ORN) in a sub-threshold concentration~-- i.e., when the
average value of the number of the ORN bound receptor proteins (RPs)
is insufficient for the generation of spikes, but such a generation
is still possible due to fluctuations around the average value~--
the ORN selectivity can be higher than the selectivity at higher
concentrations and, in particular, higher than the selectivity of
the ORN's RPs.\,\,In this work, the optimal odorant concentration
providing the highest ORN selectivity is found in the framework of a
simplified ORN model, and the dependence of the highest selectivity
on the total number of RPs in the ORN, $N$, and its threshold value
$N_{0}$ is derived.\,\,The effect of enhanced selectivity in the
sub-threshold regime is best manifested, if $N_{0}$ is close to
either unity or $N$.\,\,It is also more pronounced at large
$N$-values.\medskip

\end{abstract}

\section{Introduction}

The identification of substances in air by living organisms is
performed via the olfactory sensory system in the form of odor
reception/recognition.\,\,The olfactory system has a hierarchical
organization \cite{Ressler1994}.\,\,In particular, neurons at every
hierarchical level have a better selectivity and sensitivity to
odors than those at the previous one (see, e.g.,
\cite{Duchamp1982}).\,\,A better selectivity of secondary olfactory
neurons in comparison with primary ones is explained by the
mechanism of lateral inhibition in the olfactory bulb
\cite{Rall1968}.\,\,For low odorant concentrations, when the lateral
inhibition mechanism does not work \cite{Duchamp1982}, another
mechanism has been proposed \cite{Vidybida2019a}, which is
physically close to that considered in this~work.

The primary reception of odors and the first stages of processing
the relevant information are similar in most living organisms
\cite{Hildebrand1997}.\,\,The very first neuron that responds to the
odor is the olfactory receptor neuron (ORN).\,\,The ORN is usually
considered to be the first level in the hierarchical reception of
odors.\,\,But the reception of an odor by the ORN includes two
consecutive stages.\,\,The first stage is purely physical (see Sec\-tion~1.1%
).\,\,At some parts of its surface that are exposed to the external
environment, the ORN has a substantial number of identical receptor
proteins (RPs).\,\,Wi\-thin the same organism, there are many
different types of RPs, and there are many neurons that carry RPs of
the same type \cite{Buck2000}.\,\,Be\-cause of the Brownian motion,
odorant molecules can release the RP which they are bound with and
bind to another RP.\,\,When binding an RP, ion channels become open
in the ORN membrane.\,\,As a result, the membrane depolarizes, and
there arises a receptor potential.\,\,If the depolarization is
sufficient to excite and generate output impulses, the ORN sends
them to secondary neurons.

A separate ORN reacts differently to different odorants (it sends
impulses with different frequencies).\,\,In addition, ORNs with
different RPs react differently to the same odorant.\,\,This
circumstance makes it possible to create a combinatorial code that
allows the distinguishing of many more odors than the number of
different RP types~\cite{Malnic1999}.

Earlier, it was predicted theoretically \cite{Vidybida2022} that if
odorants are applied to the ORN at concentrations lower than it is
required for a stable generation of spikes (sub-threshold ones) so
that only their random generation due to fluctuations is possible,
the ORN selectivity can be substantially enhanced.\,\,In this paper,
possible ORN parameters and concentrations that provide the maximum
enhancement of selectivity are estimated.\,\,At the same time, an
extremely simple ORN model is used, which takes into account only
the statistics of the binding-release process of odorant molecules
by receptor proteins.\,\,The\-refore, the obtained results do not
pretend to be an adequate description of the phenomena in the
biological ORN.\,\,They can be interpreted only as a hint of what
parameter values could improve the selectivity as much as
possible.\,\,The estimates made here can be used for setting up
experiments with real neurons and under conditions of low odorant
concentrations in order to provide the maximum selectivity, as well
as for designing artificial \mbox{chemosensors}.

\subsection{Primary reception of odors}

\label{PSZ}

From the physical point of view, the primary reception of an odorant
molecule in the olfactory system occurs in the course of the
association-dissociation of this molecule with the receptor
protein.\,\,In most cases, the
association-dissociation reaction runs according to the following scheme:%
\def\chh#1#2{\vcenter{\vbox{\hbox{ #1\vspace*{-1mm} }\hbox{ $\rightleftharpoons$
\vspace*{-1mm}}\hbox{ #2 } } }}
\begin{equation}\label{chem1}
                 \mathbf{O}+R \chh{\footnotesize $k_+$}{\footnotesize $k_-$} \mathbf{O}R,
\end{equation}
where $\mathbf{O}$ is the odorant molecule, and $R$ is the receptor
protein.\,\,This is the first step in the odor reception
process.\,\,It results in that some of receptor proteins will be
occupied by the odorant molecules, whereas the remaining RPs will
remain free.\,\,The quantitative measure of this result is the ratio
\begin{equation}
p=\frac{n}{N}\label{p}%
\end{equation}
between the number of occupied RPs, $n$, and the total RP number,
$N$, in a single ORN.\,\,If, in the course of two independent
experiments, an ORN receives two different odorants with the same
concentration, but the fractions of occupied RPs are different, then
the RP can distinguish between those two odors, i.e., it is
selective with respect to them.\,\,If the fractions are equal, the
RP is not able to do this.\,\,In the latter case, the corresponding
ORN will also be not able to distinguish between the indicated two
odorants, because the depolarizing transmembrane current, which
governs the rate of spike generation by the neuron, depends on the
number of occupied RPs.

\subsubsection*{1.1.1.\,\,ORN selectivity}

\begin{wrapfigure}[18]{R}[-2mm]{0.5\textwidth}
\centering
\vskip1mm
\includegraphics[scale=0.95]{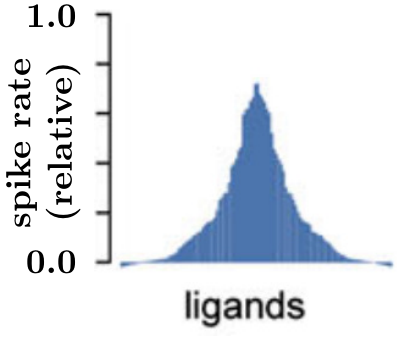}
\vskip-3mm
\caption{Example of different frequencies of output spikes
generated by an ORN for one type of RPs stimulated by different
odorants. Adapted from \cite{Galizia2010} (through Creative Commons
public license, {https://creativecommons.org}). See also
\cite[Fig.~3]{Hildebrand1997}  }\label{bell}
\end{wrapfigure}
The ultimate result of the odor reception by an olfactory receptor
neuron is the generation of output impulses by this neuron.\,\,Most
ORNs generate impulses as a response to lots of various odors: they
are generalists rather than specialists.\,\,The ability of ORNs to
distinguish between two odorants manifests itself in the different
impulse frequencies, if the odorants are presented in the same
concentration in two independent experiments.\,\,For a set of
odorants that a single ORN reacts to, a curve that conditionally
characterizes the ORN selectivity can be plotted (see Fig.~1).

We now put a question: Is the selectivity of ORN identical to that
of its receptor proteins? It is clear that the larger the share of
RPs bound to odorant molecules, the larger the depolarization of the
ORN excitatory membrane and the higher the generation frequency of
output spikes by the ORN.\,\,This relation connects the selectivity
of the ORN with that of its RP.\,\,Ho\-we\-ver, in view of
complicated intermediate mechanisms of chemo-electrical transduction
from the RP binding-release to the creation of receptor potential
and further to the spike generation, there is no reason to equate
the selectivities of RPs and ORN expressing those proteins.

The aim of this work is to elucidate at which ORN parameters and
odorant concentrations one may expect the highest selectivity of the
ORN assuming the random binding-release of its RPs.\,\,For this
purpose, the simplest ORN model was applied, where all intermediate
stages of chemo-electrical transduction giving rise to the spike
generation are replaced by the fact of reaching the threshold value
by the number of bound RPs.\,\,The reception regime in which
fluctuations in the number of bound RPs substantially affect the
spike generation, the sub-threshold regime, is also considered
(Section~2.1.1).\,\,Pre\-vious results obtained in the framework of
this model \cite{Vidybida2022} showed that it is possible to obtain
the selectivity of ORN in the sub-threshold regime that considerably
exceeds the selectivity of its RPs.

\section{Methods}

\subsection{Membrane-free ORN model}

\label{BM}

The ORN model analyzed here includes only the events that happen at
the outer ORN surface in the course of the interaction between the
ORN's RPs and odorant molecules.\,\,This model is similar to that
discussed in work~\cite{Lamine1997}, but is even simpler, because it
does not consider the passage of odorant molecules through the mucus
of olfactory epithelium.\,\,In the framework of this model, the ORN
is characterized by the total number $N$ of identical receptor
proteins incorporated into its membrane, and the threshold number
$N_{0}<N$.\,\,If the number of bound RPs is less than $N_{0}$
($n<N_{0}$), then the ORN does not generate spikes.\,\,In the
opposite case, the ORN generates spikes at a constant frequency $f$
(cf.~\cite[Section \textquotedblleft Olfactory
threshold\textquotedblright]{Lamine1997}).

It should be noted that the application of this model implies that
the binding of one RP with an odorant molecule opens one ion
channel.\,\,This is the case for the ORN of insects, where the
receptor proteins are heteromeric ligand-gated ion
channels~\cite{Sato2008}.\,\,In the ORN of more complicated
organisms, intermediate biochemical events take place between the RP
binding and the opening of ion channels, and, as a result, the
binding of one RP provides the opening of several channels, which
are structurally separated from the RP (see,
e.g.,~\cite{Ronnett2002}).\,\,Those intermediate events are an
additional source of fluctuations and require the additional
analysis in a sepa\-rate~\mbox{paper.}\looseness=1

\subsubsection*{2.1.1.\,\,Sub-threshold regime}

\label{ppr}To simplify calculations, it is assumed here that the ORN
generates output impulses at a constant frequency $f$ irrespective
of how much the threshold $N_{0}$ is exceeded.\,\,This assumption is
a substantial deviation from reality, if the odorant concentration
is high, and the number of bound RPs, $n$, permanently exceeds the
threshold value $N_{0}$.\,\,In this case, the growth of $n$
increases the frequency of output impulses.\,\,But, in this work,
the consideration is focused on low concentrations, when the average
number of bound RPs is less than the threshold value, and the
threshold is reached for short time intervals due to fluctuations
(see Section~2.3).\,\,It is assumed that either one or no impulse
can be generated during the permanent stay above the
threshold.\,\,In this regime, the average frequency of output
impulses is governed by the probabilistic characteristics of the
threshold crossing, rather than the degree of threshold exceedance.

In order to strictly substantiate that the sub-threshold regime
described above is possible, it is necessary to know the temporal
characteristics of the stochastic process of RP binding-release and
the kinetics of the process of generating output impulses by the
excitable neuronal membrane.\,\,Those parameters include the
reaction rate constants, the conductivity of the channels that
become open at the RP binding, and the electrical characteristics of
the membrane.\,\,Those parameters can be taken into account in
numerical simulations.\,\,In this work, we do not specify them and
intend to do so in the future.

\subsection{Definition of selectivities}

The selectivity of RP and ORN can be defined in various
ways.\,\,Here, we follow the definitions from
work~\cite{Vidybida2022}.\,\,They exclude the consideration of the
concentration and the dissociation reaction constant~$K$ [see
Eq.~(\ref{chem1}) below], and the consideration is based on the
fraction $p$ of bound RPs~(\ref{p}).\,\,This approach is justified
for two reasons.\,\,First, it is simpler to deal with
$p$.\,\,Be\-sides, formula~(\ref{pK}) given below provides an
unambiguous relationship between $K$ and $p$, if the concentration
$c$ is fixed, or between $c$ and $p$ if the dissociation constant
$K$ is fixed.\,\,Se\-cond, the olfactory neuron has no access to the
$K$- and $c$-values, whereas the information about the total RP
number $N$ and the number $n$ of bound RPs on the neuron surface
[which, according to formula (\ref{p}), is equivalent to knowing the
$p$-value] is exactly what is subjected to a further processing in
the ORN and invokes the generation of output impulses.

The selectivity of RPs with respect to two odorants $\mathbf{O}_{1}$
and
$\mathbf{O}_{2}$ is defined as follows.\,\,If $\mathbf{O}_{1}$ and $\mathbf{O}%
_{2}$ are presented to the ORN at the same concentration $c$ in two
independent experiments, and if different $p$-values, $p_{1}$ and
$p_{2}$, are observed at that, then this RP can distinguish between
those two odorants.\,\,For definiteness, let $p_{1}>p_{2}$, i.e.,
\begin{equation}
p_{1}=p_{2}+\Delta p,\quad\Delta p>0. \label{ineqp}%
\end{equation}
Then the RP selectivity can be defined as follows:
\begin{equation}
S_{R}=\frac{\Delta p}{p_{1}}. \label{SR}%
\end{equation}

For the entire ORN, its reaction to the odor manifests itself as the
generation of output impulses.\,\,We may expect that, owing to
Eq.~(\ref{ineqp}), the average impulse frequency $F$ will be higher
for $\mathbf{O}_{1}$, i.e.,
\begin{equation}
F_{1}=F_{2}+\Delta F,\quad\Delta F>0. \label{ineqF}%
\end{equation}
Then the ORN selectivity can be defined as follows:
\begin{equation}
S_{\mathrm{ORN}}=\frac{\Delta F}{F_{1}}. \label{SORN}%
\end{equation}

By analogy with \cite{Lamine1997}, if we assume that, at high
odorant concentrations, when the number of occupied RPs permanently
exceeds the excitation threshold, the ORN response is proportional
to the number of occupied RPs, then the ORN selectivity will be
equal to the selectivity of its RPs.\,\,In\-deed, in our case, the
ORN response is the average impulse frequency $F$.\,\,If $F$ grows
proportionally with $n$, then
\begin{equation}
S_{\mathrm{ORN}}=\frac{N\Delta p}{Np_{1}}=\frac{\Delta p}{p_{1}}=S_{R}.
\label{SORNH}%
\end{equation}
Therefore, for concentrations providing a permanent exceedance of the
excitation threshold, the ORN selectivity in the simple transduction model is
identical to the selectivity of its RPs.

If the odorant concentration is sub-threshold, and if the $N_{0}$
threshold is exceeded due to fluctuations during short time
intervals, then the ORN response will be determined by the fraction
of  time  the number of bound RPs spends above the excitation
threshold.\,\,Be\-low, we analyze how the differences between the
statistics of random threshold crossings for the $\mathbf{O}_{1}$
and $\mathbf{O}_{2}$ odorants determine the ORN selectivity.

\subsection{Primary-reception fluctuations}

\label{flu}

Since the primary reception of odor by a receptor neuron is
performed through the binding and release of odorant molecules by
the neuron's receptor proteins, this event is inevitably
random.\,\,As a result, secondary signals about the odor, such as
the membrane (receptor) potential or the transmembrane current, will
also be random.\,\,Fluc\-tua\-tions of the transmembrane current in
the ORN of the amphibian \textit{Ambystoma tigrinum} were observed
experimentally \cite{Menini1995}; minimum odorant concentrations,
10$^{-10}$$\div$$5\, \times$ $\times\, 10^{-7}$$\div$$10^{-5}$M,
were used at that.\,\,The olfactory receptor neurons of amphibians
have a more complicated mechanism of chemo-electrical transduction
than in the case of insects (see, e.g., \cite{Ronnett2002}); in
particular, it allows the temporal integration of weak stimuli
\cite{Menini1995}.\,\,In this work, in the framework of the
simplified ORN model, we do not consider the possibility of the
temporal integration.

When an odorant \textbf{O} is applied to an ORN, the RPs of the
latter, due to the Brownian motion, randomly bind \textbf{O}
molecules and get released from them.\,\,Here, it is assumed that
the random behavior of a separate RP is independent of other
RPs.\,\,Af\-ter the completion of transient processes, every RP
belonging to a certain ORN can be bound to an \textbf{O} molecule
with a certain probability.\,\,Note that this probability is equal
to $p$ defined in Eq.~(\ref{p}).\,\,If the concentration $c$ of the
applied odorant \textbf{O} and the dissociation constant $K$ for the
association-dissociation reaction (\ref{chem1}) between \textbf{O}
and RP are known, then, according to the
known formula (cf.\,\,\cite[Eq.\,\,(3)]{Chastrette1998} and \cite[Eq.\,\,(4)]%
{Lamine1997}),\vspace*{-1mm}
\begin{equation}
p=\frac{1}{1+K/c}.\label{pK}%
\end{equation}

For the model described in Section~2.1, it is important to know the
probability $\mathbf{P}$ of that the number of bound RPs exceeds the
threshold value $N_{0}$ or reaches it provided that the odorant
applied to the ORN ensures a certain fraction $p$ (on average) of
occupied RPs.\,\,Since, as was indicated above, this fraction is
also the probability of that a single RP is bound to an odorant
molecule, then, if separate RPs are statistically independent, the
sought probability $\mathbf{P}$ of reaching/exceeding the threshold
can be calculated using the known formula (see, e.g.,
\cite[Chap.\,\,3, Eq.\,\,(1)]{GnedenkoUKR1950})\vspace*{-1mm}
\begin{equation}\label{PNN0p}
\mathbf{P}(N,N_0,p)=
\sum\limits_{k=N_0}^N\binom{\!N}{\!k}p^k(1-p)^{N-k}.
\end{equation}

In\,\,the\,\,described\,approach,\,the\,quantity\,$\mathbf{P}(N,N_{0},p)$
is the probability of that the threshold will be reached/exceeded at
some time moment, and the frequency $f$, which was introduced in
Section~2.1, is a dimensional multiplier, which makes it possible to
calculate the average frequency of output impulses,\vspace*{-1mm}
\begin{equation}
F=f\,\mathbf{P}(N,N_{0},p). \label{fPF}%
\end{equation}
The value of $f$ is nonessential for the definition of selectivity
(\ref{SORN}),
\begin{equation}
S_{\mathrm{ORN}}=\frac{\mathbf{P}(N,N_{0},p_{1})-\mathbf{P}(N,N_{0},p_{2}%
)}{\mathbf{P}(N,N_{0},p_{1})}. \label{SORNP}%
\end{equation}

\section{Results}

\subsection{Optimal concentration}

\label{OK}

\begin{wrapfigure}[19]{R}[-2mm]{0.5\textwidth}
\centering
\vskip1mm
\includegraphics[width=0.40\textwidth,angle=-0]{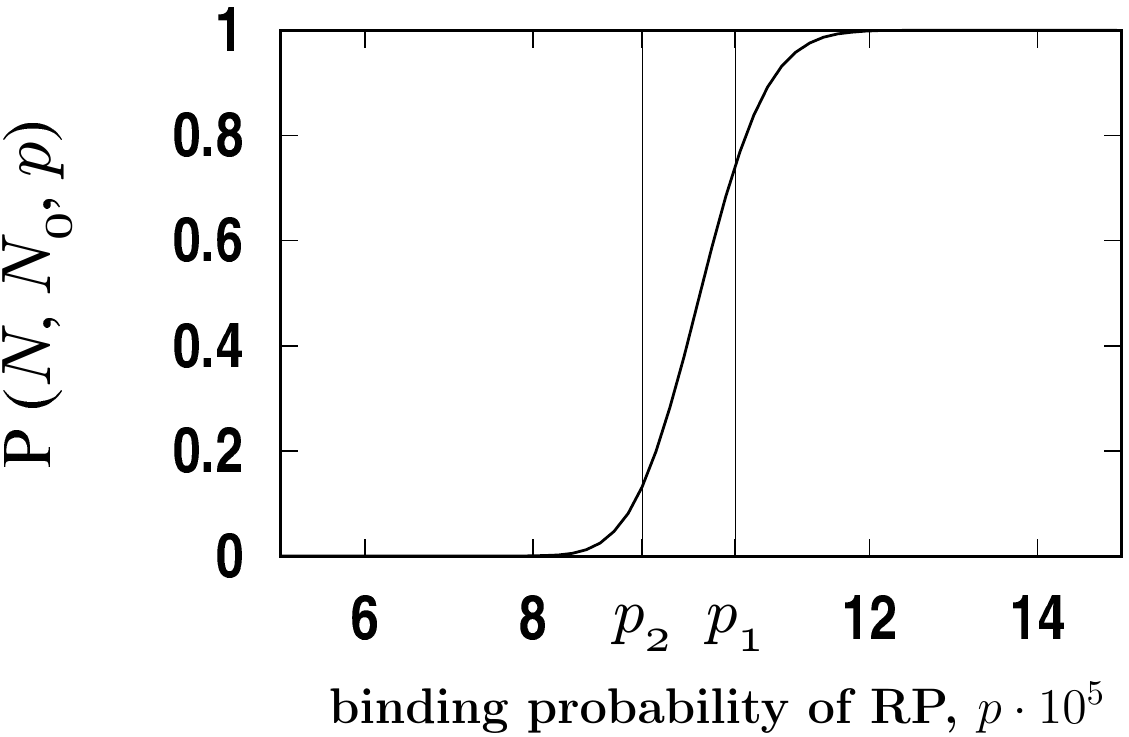}
\vskip-3mm\caption{The probability of reaching the threshold,
$\mathbf{P}(N,N_{0},p)$, for $N=2\,500\,000$ and $N_{0}=250$. Here,
$p_{2}=0.9296\times10^{-4}$, $p_{1}=1.040\times10^{-4}$, $S_{R}=0.1$
[see Eq.~(\ref{SR})], and $S_{\mathrm{ORN}}=0.8$ [see
Eqs.~(\ref{SORN}) and (\ref{SORNP})]. The values for $N$ and $N_{0}$
were approximately chosen on the basis of the data for the moth
\emph{Antheraea polyphemus} from work~\cite{Kaissling2001}
}\label{sigmoid}
\end{wrapfigure}
Expression~(\ref{PNN0p}) for $\mathbf{P}(N,N_{0},p)$ and expression~(\ref{fPF}%
) for $F$ depend on $p$ in a \textquotedblleft sigmoid\textquotedblright%
\ manner, i.e., they first grow slowly, then enter the interval with
a rapid growth, and afterward slowly saturates to the corresponding
constant value.\,\,Ta\-king into account that $p$ increases
monotonically with $c$ [see Eq.~(\ref{pK})], the dependences of
$\mathbf{P}(N,N_{0},p)$ and $F$ on $c$ will also be qualitatively
the same.\,\,If the odorants $\mathbf{O}_{1}$ and $\mathbf{O}_{2}$
have almost the same affinity to the RP, then the corresponding
values of $p_{1}$ and $p_{2}$ will be very close to each other,
which means a low selectivity of RP with respect to those
odors.\,\,If the concentration of odorants is such that $p_{1}$ and
$p_{2}$ fall into the interval of a rapid $\mathbf{P}(N,N_{0},p)$
growth, one may expect a large difference between the average
frequencies of ORN impulses for those two odors [see
Eq.~(\ref{fPF})].\,\,This will mean a better ORN
selectivity.\,\,This situation is illustrated in Fig.~2.

To determine the optimal values of $p$ and $c$, we have to find the
point $p_{0},$ where the $p$-derivative of $\mathbf{P}(N,N_{0},p)$
is maximum.\,\,This derivative equals
\begin{equation}\label{1st}
\frac{d}{dp}\mathbf{P}(N,N_0,p)
  =\frac{N!}{(N_0-1)!(N-N_0)!}p^{N_0-1}(1-p)^{N-N_0}.
\end{equation}
To find its maximum, expression~(\ref{1st}) has to be differentiated
once more,
\begin{equation}\label{2nd}
\frac{d^2}{dp^2}\,\mathbf{P}(N,N_0,p) 
  \sim p^{N_0-2}(1-p)^{N-N_0-1} (N_0-1 - p(N-1))=0.
\end{equation}
(here, the multiplier independent of $p$ is omitted).\,\,From
Eq.~(\ref{2nd}), we have
\begin{equation}
p_{0}=\frac{N_{0}-1}{N-1}. \label{p0}%
\end{equation}
Therefore, the optimal concentration $c_{0}$ should provide the
average number of bound RPs that is below $N_{0}$ and above
$N_{0}-1$.\,\,The corresponding $c_{0}$-value is obtained from
Eqs.~(\ref{pK}) and~(\ref{p0}),
\begin{equation}
c_{0}=\frac{K(N_{0}-1)}{N-N_{0}}. \label{c0}%
\end{equation}

This work is not aimed at elucidating the possible mechanisms for
creating the exact optimal concentration (however, see
Section~4).\,\,But it is clear that the effect of enhanced
selectivity will be observed in a certain interval of $p$-values
around $p_{0}$, which is illustrated in Fig.~2.

\subsection{Influence of threshold magnitude}

\begin{wrapfigure}[16]{L}[-2mm]{0.5\textwidth}
\vskip1mm
\includegraphics[height=0.45\textwidth,angle=-90]{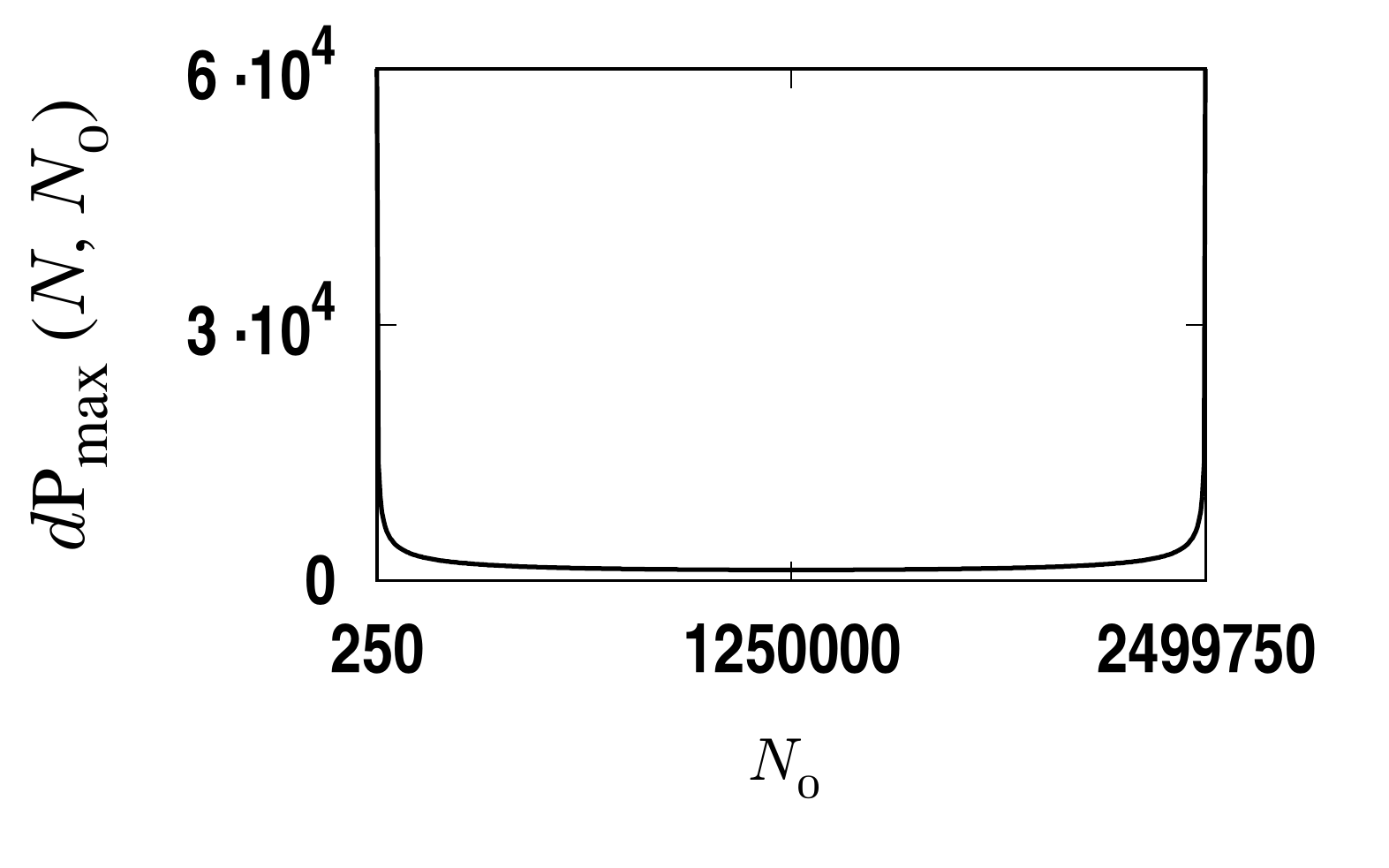}
\vskip-3mm\caption{Dependence of the maximum value of the derivative $d\mathbf{P}_{\max}(N,N_{0})$ on the threshold magnitude $N_{0}$ for the fixed
total number
of RPs $N=2\,500\,000$. The minimum of the function $d\mathbf{P}_{\max}%
(N,N_{0})$ equals 1262. The function values at the points
$N_{0}=250$ and $N_{0}=N-250$ are approximately identical and
approximately equal to $63\times10^{3}$ }\label{PmN0}
\end{wrapfigure}
In the previous section, it was found what concentration of weakly
different odorants should be for the best manifestation of the
effect of ORN selectivity enhancement in comparison with that of its
RPs, if the total number of RPs in the neuron, $N$, and the
threshold value $N_{0}$ are fixed.\,\,The optimal
concentration provides the optimal binding probability $p_{0}$ [Eq.~(\ref{p0}%
)], such that the derivative of $\mathbf{P}(N,N_{0},p)$ with respect
to $p$ is largest at the point $p_{0}$.\,\,But, the manifestation of
the selectivity enhancement effect depends on the absolute value of
the derivative at the point $p_{0}$.\,\,This value is determined by
the quantities $N$ and $N_{0}$.
Let us elucidate how the maximum value of the derivative, $d\mathbf{P}%
_{\max}(N,N_{0})$, depends on $N_{0}$ at a fixed $N$.\,\,For this
purpose, let us substitute $p$ by $p_{0}$ in
formula~(\ref{1st}).\,\,We obtain
\begin{equation}
\label{1stmN}
d\mathbf{P}_{max}(N,N_0)=
\frac{d}{dp}\mathbf{P}(N,N_0,p)\bigg|_{p=p_0}=
N\,\binom{N-1}{N_0-1}p_0^{N_0-1}(1-p_0)^{N-N_0}\,,
\end{equation}

\noindent where $p_{0}$ is given in (\ref{p0}).\,\,If $N_{0}=1$,
formula (\ref{p0}) gives $p_{0}=0$.\,\,It is clear that the
selectivity to odorants with zero concentration has no
sense.\,\,But, the value of $d\mathbf{P}_{\max}(N,1)$ can give an
estimate of the slope of the plot of the function
$\mathbf{P}(N,1,p)$ in a vicinity of the point $p=0$, and this may
be interesting in the case of a very low
concentration\,\footnote{Note that the ability of mice to detect
some odorants at a concentration of $10^{-11}$ M was observed
experimentally \cite{Williams2020}.\,\,The authors of work
\cite{Bhandawat2010} gave a value of $10^{-13}$ M for the
theoretical estimate of the minimum concentration that can be
detected by the olfactory system.}.\,\,The required value can be
found as the limit\vspace*{-2mm}
\begin{eqnarray}\nonumber
  d\mathbf{P}_{\max}(N,1)=
\\\nonumber
  = \lim\limits_{p\to0}
{N}\!\binom{\!N-1}{0}\!p^{0}(1-p)^{N-1}=
\\\label{1stmN01}
=N.
\end{eqnarray}

For $N_{0}=2$, we have $d\mathbf{P}_{\max}(N,2)\approx\frac{N}{e}$.
For large $N$ and $N_{0}$, by applying the Stirling formula to
Eq.~(\ref{1stmN}), we obtain the approximate value\vspace*{-1mm}
\begin{equation}
d\mathbf{P}_{\max}(N,N_{0})\approx N\,\sqrt{\frac{N-1}{2\pi(N_{0}%
-1)(N-N_{0})}}\,. \label{Stirl}%
\end{equation}
From whence, we can see that $d\mathbf{P}_{\max}(N,N_{0})$
increases, as $N$, which is in agreement with
formula~(\ref{1stmN01}).\,\,An example of the plot for
$d\mathbf{P}_{\max}(N,N_{0})$ is shown in Fig.~3.

\subsection{Illustrative example}

\begin{wrapfigure}[26]{R}[-2mm]{0.5\textwidth}
\vskip3mm
\vbox{\hbox{\includegraphics[height=0.40\textwidth,angle=-90]{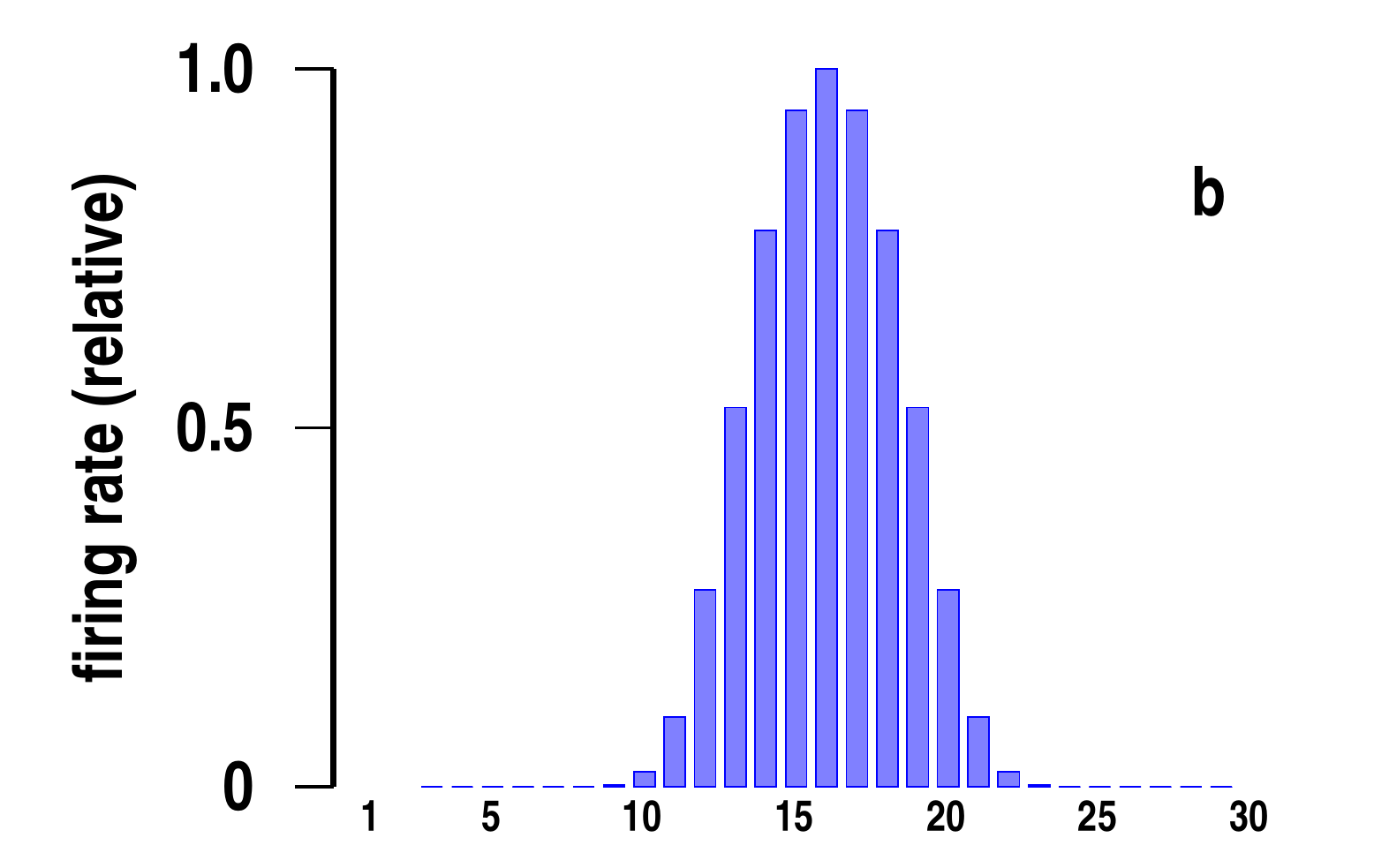}}%
\hbox{\includegraphics[height=0.40\textwidth,angle=-90]{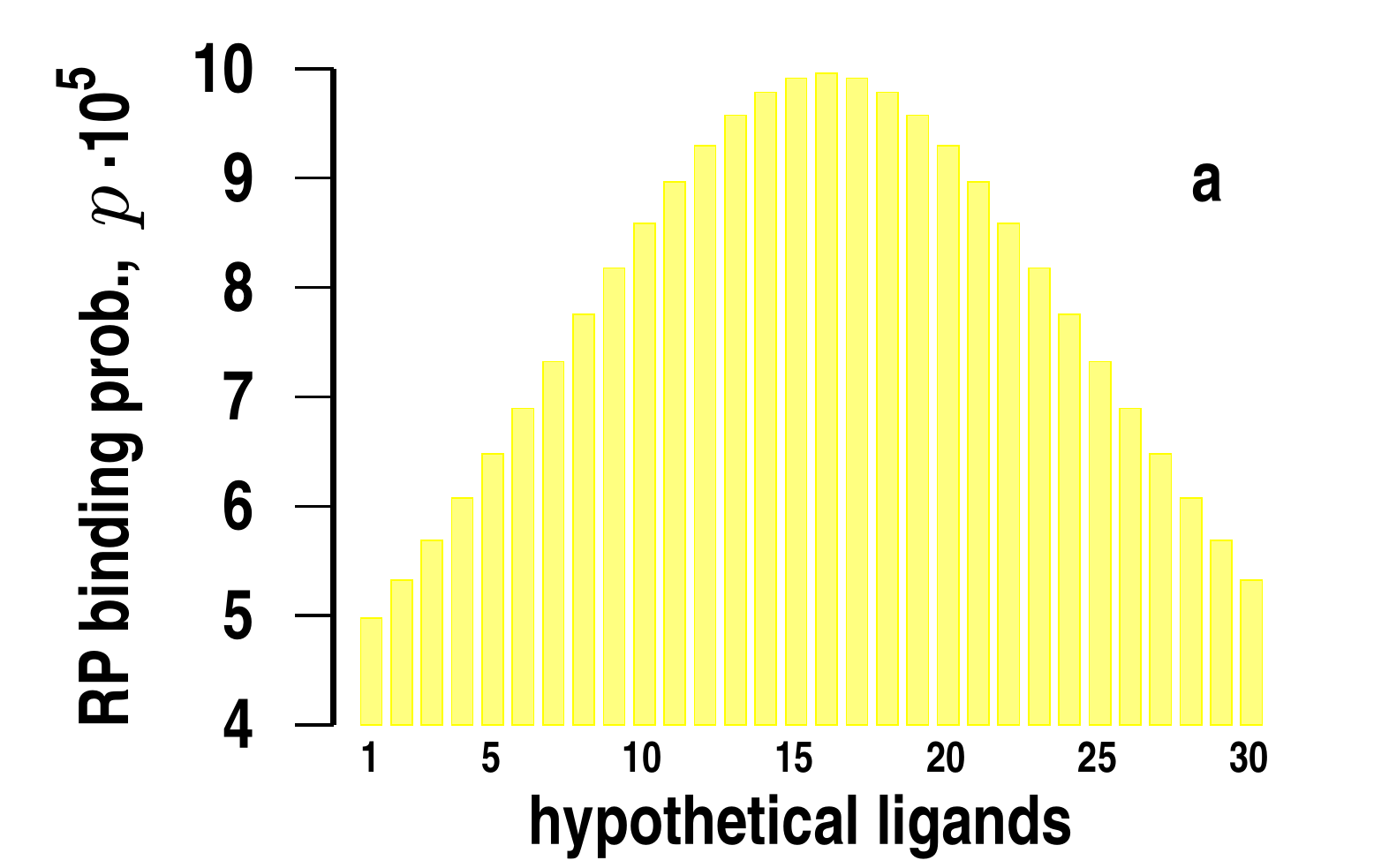}}}
 \vskip-3mm\caption{Illustration that in the sub-threshold
regime, an ORN can possess a better selectivity than its RPs:
fraction $p$ of bound RPs for a set of 30 hypothetical odors with
various affinities with respect to RPs ($a$); relative frequency of
the spike generation by the ORN, when the indicated hypothetical
odorants are applied  ($b$)}\label{ipF}
\end{wrapfigure}
To compare the selectivity of ORN with that of its RPs, selectivity
plots similar to the plot shown in Fig.~1 were drawn.\,\,For this
purpose, a set of 30 different $p$-values inherent to hypothetical
odors were generated.\,\,The obtained RP selectivity plot has a wide
bell-shaped form (Fig.~4,~$a$).\,\,To obtain the relative
frequencies of ORN spikes (Fig.~4,~$b$), those 30 indicated
$p$-values were used in formulas (\ref{PNN0p}) and (\ref{fPF}).

The selectivities between odors \#9 and \#16 from Fig.~4, which were
calculated according to formulas (\ref{SR}) and (\ref{SORN}),
acquire the values\vspace*{-1mm}
\begin{equation}
S_{R}=0.178,\quad S_{\mathrm{ORN}}=0.998. \label{Srporn}%
\end{equation}\vspace*{-5mm}
\newpage

\section{Discussion and Conclusions}

\label{VD}

In this work in the framework of the simplified model for an
olfactory receptor neuron, two conditions are found that provide the
maximum enhancement of the ORN selectivity as compared to that of
the ORN receptor proteins.\,\,The first condition is the
sub-threshold regime of odor reception.\,\,It is provided by
selecting the odorant concentration [Eqs.~(\ref{p0}) and
(\ref{c0})]. The second condition is the minimum number $N_{0}$ of
bound receptor proteins required for the ORN to start the spike
generation [Eqs.~(\ref{1stmN}) and (\ref{Stirl})].

From Fig.~3, it is seen that the selectivity enhancement in the
sub-threshold regime is larger for ORNs with lower triggering
thresholds and for very low concentrations.\,\,For real ORNs, these
conditions can be satisfied only partially.\,\,First, the threshold
magnitude $N_{0}$ is dictated by the electrical properties of the
ORN membrane and the ion channels connected with every RP.\,\,The
minimum values of $N_{0}$ measured for the frog ORNs are about $35$
\cite{Bhandawat2010}.\,\,But, each bound RP in the frog ORN opens
several ion channels by means of the mechanism described in work
\cite{Ronnett2002}.\,\,For insects, where one RP opens one channel,
a threshold value of several hundreds seems to be close \mbox{to
reality.} 
Second, the total number of RPs in an ORN has to be large [see
Eqs.~(\ref{1stmN})--(\ref{Stirl})].\,\,But, is it possible to affect
the value of $N$ fast enough? The first condition governs the way
that the odor is presented, whereas the second one is responsible
for the ORN structure or dynamic characteristics.

The biological olfactory system has the means to satisfy those
conditions within certain limits.\,\,First, air with the dissolved
odorant does not contact directly with the ORN surface, but through
the mucus.\,\,The latter contains enzymes that chemically decompose
the odorant molecules~\cite{Nagashima2010} and control the effective
odorant concentration at the ORN surface.\,\,If the decomposition
process takes place, the respiration rate also affects the effective
concentration.\,\,Se\-cond, the level of threshold depolarization of
the excitable ORN membrane depends on the ionic composition of the
environment near the membrane.\,\,Changing this composition, we can
affect $N_{0}$.\,\,Third, some biological
mechanisms~\cite{Bryche2021}, with the RP internalization among
them, can affect the number $N$ of RPs at the ORN surface.

The conditions above can be satisfied in artificial neuromorphic
sensors like biosensors or the electronic nose
\cite{Persaud1982,Raman2011,Hurot2020,Kucherenko2020,Kim2021}.\,\,For
such devices, the case of very high concentrations would also be of
interest.\,\,As one can see from Fig.~3 (the right-hand side of the
plot corresponding to large $N_{0}$-values), if the concentration is
close to the saturation, the quantity $\mathbf{P}(N,N_{0},p)$
regarded as a function of $p$ also changes very quickly in a
vicinity of $p_{0}$.\,\,Ho\-we\-ver, the accurate registration of
threshold crossing in the case where the threshold magnitude is
equal to several millions will be problematic.\,\,On the other hand,
the artificial sensor is capable of detecting the number of free
receptors, which is small at high concentrations.

Some deviations of the considered model from the real ORN have been
specified above.\,\,It is worth adding here that real neurons vary
in time.\,\,If an ORN is subjected to a permanent exposure to the
odor, its sensitivity decreases, and the adaptation phenomenon is
observed~\cite{Antunes2014}.\,\,The spontaneous activity of ORN in
the absence of odorants~\cite{JobyJoseph2012} was also not
considered here.\,\,Be\-sides, we note that the analysis of the
fluctuations of the primary response in chemical sensors is also
applied beyond the receptor binding-release
statistics~\cite{SMULKO2001, Scandurra2020}.

\vskip3mm {\it This work was sponsored in the framework of the
Fundamental research program of the Branch of Physics and Astronomy
of the National Academy of Sciences of Ukraine (project
No.\,0120U101347 \textquotedblleft Noise-induced dynamics and
correlations in non-equilibrium systems\textquotedblright) and by
the private American Simons Foundation.}


\begin{thebibliography}{99}                                                                                               %


\bibitem {Ressler1994}K.J. Ressler, S.L. Sullivan, L.B. Buck. Information
coding in the olfactory system: evidence for a stereotyped and highly
organized epitope map in the olfactory bulb. \emph{Cell} \textbf{79}, 1245
(1994).\vspace{0.3mm} 

\bibitem {Duchamp1982}A.~Duchamp. Electrophysiological responses of olfactory
bulb neurons to odour stimuli in the frog. a comparison with receptor cells.
\emph{Chem. Sens.} \textbf{7}, 191 (1982).\vspace{0.3mm} 

\bibitem {Rall1968}W.~Rall, G.M. Shepherd. Theoretical reconstruction of field
potentials and dendrodendritic synaptic interactions in olfactory bulb.
\emph{J, Neurophysiol.} \textbf{31}, 884 (1968).\vspace{0.3mm} 

\bibitem {Vidybida2019a}A.K. Vidybida. Possible stochastic mechanism for
improving the selectivity of olfactory projection neurons.
\emph{Neurophysiology} \textbf{51}, 152 (2019).\vspace{0.3mm} 

\bibitem {Hildebrand1997}J.G. Hildebrand, G.M. Shepherd. Mechanisms of
olfactory discrimination: Converging evidence for common principles across
phyla. \emph{Annu. Rev. Neurosci.} \textbf{20}, 595 (1997).\vspace{0.3mm} 

\bibitem {Buck2000}L.B. Buck. The molecular architecture of odor and phe\-romone
sensing in mammals. \emph{Cell} \textbf{100}, 611 (2000).\vspace{0.3mm} 

\bibitem {Malnic1999}B.~Malnic, J.~Hirono, T.~Sato, L.B. Buck. Combinatorial
receptor codes for odors. \emph{Cell} \textbf{96}, 713 (1999).\vspace{0.3mm} 

\bibitem {Vidybida2022}A.~Vidybida. Harnessing thermal fluctuations for
selectivity gain. In \emph{2022 IEEE International Symposium on Olfaction and
Electronic Nose (ISOEN) }(2022), p.~1. 

\bibitem {Galizia2010}C.G. Galizia, D.~Munch, M.~Strauch, A.~Nissler, Shouwen
Ma. Integrating heterogeneous odor response data into a common response model:
A door to the complete olfactome. \emph{Chem. Sens.} \textbf{35}, 551 (2010). 

\bibitem {Lamine1997}A.B. Lamine, Y.~Bouazra. Application of statistical
thermodynamics to the olfaction mechanism. \emph{Chem. Sens.} \textbf{22}, 67
(1997).

\bibitem {Sato2008}K.~Sato, M.~Pellegrino, T.~Nakagawa, T.~Nakagawa,
L.B.~Voss\-hall, K.~Touhara. Insect olfactory receptors are
he\-te\-ro\-meric ligand-gated
ion channels. \emph{Nature} \textbf{452}, 1002 (2008). 

\bibitem {Ronnett2002}G.V. Ronnett, Ch. Moon. \textit{G} proteins and
olfactory signal transduction. \emph{Annu. Rev. Physiol.} \textbf{64}, 189
(2002). 

\bibitem {Menini1995}A.~Menini, C.~Picco, S.~Firestein. Quantal-like current
fluctuations induced by odorants in olfactory receptor cells. \emph{Nature}
\textbf{373}, 435 (1995). 

\bibitem {Chastrette1998}M.~Chastrette, T.~Thomas-Danguin, E.~Rallet.
Modelling the human olfactory stimulus-response function. \emph{Chem. Sens.}
\textbf{23}, 181 (1998). 

\bibitem {GnedenkoUKR1950}B.V. Gnedenko. \textit{Theory of Probability. 6th
Edition }(CRC Press, 1998).

\bibitem {Kaissling2001}K-E Kaissling. Olfactory perireceptor and receptor
events in moths: A kinetic model. \emph{Chem. Sens.} \textbf{26}, 125 (2001). 

\bibitem {Williams2020}E.~Williams, A.~Dewan. Olfactory detection thresholds
for primary aliphatic alcohols in mice. \emph{Chem. Sens.} \textbf{45}, 513
(2020). 

\bibitem {Bhandawat2010}V.~Bhandawat, J.~Reisert, K-W Yau. Signaling by
olfactory receptor neurons near threshold. \emph{Proc. Nat. Acad. Sci. USA
}\textbf{107}, 18682 (2010). 

\bibitem {Nagashima2010}A.~Nagashima, K.~Touhara. Enzymatic conversion of
odorants in nasal mucus affects olfactory glomerular activation patterns and
odor reception. \emph{J. Neurosci.} \textbf{30}, 16391 (2010). 

\bibitem {Bryche2021}B.~Bryche, C.~Baly, N.~Meunier. Modulation of olfactory
signal detection in the olfactory epithelium: focus on the internal and
external environment, and the emerging role of the immune system. \emph{Cell
Tissue Res.} \textbf{384}, 589 (2021). 

\bibitem {Persaud1982}K.~Persaud, G.~Dodd. Analysis of discrimination
mechanisms in the mammalian olfactory system using a model nose. \emph{Nature}
\textbf{299}, 352 (1982). 

\bibitem {Raman2011}B.~Raman, M.~Stopfer, S.~Semancik. Mimicking biological
design and computing principles in artificial olfaction. \emph{ACS Chem.
Neurosci.} \textbf{2}, 487 (2011). 

\bibitem {Hurot2020}C.~Hurot, N.~Scaramozzino, A.~Buhot, Y.~Hou. Bio-inspired
strategies for improving the selectivity and sensitivity of artificial noses:
A review. \emph{Sensors} \textbf{20}, 1803 (2020). 

\bibitem {Kucherenko2020}I.S. Kucherenko, O.O. Soldatkin, S.V.~Dzyadevych,
A.P.~Soldatkin. Electrochemical biosensors based on multienzyme
systems: Main groups, advantages and limitations -- a review.
\emph{Analyt. Chim. Acta}
\textbf{1111}, 114 (2020). 

\bibitem {Kim2021}S.~Kim, R.~Lee, D.~Kwon, T-H. Kim, T.J. Park, \mbox{S-J.~Choi},
H-S. Mo, D.H. Kim, B-G. Park. Multiplexed silicon nanowire tunnel fet-based
biosensors with optimized multi-sensing currents. \emph{IEEE Sensor. J.}
\textbf{21}, 8839 (2021). 

\bibitem {Antunes2014}G.~Antunes, A.M. Sebasti\~{a}o, F.M.S.~de~Souza. Mechanisms
of regulation of olfactory transduction and adaptation in the olfactory
cilium. \emph{PLOS Comput. Biol.} \textbf{9}, e105531 (2014). 

\bibitem {JobyJoseph2012}J.~Joseph, F.A. Dunn, M.~Stopfer. Spontaneous
olfactory receptor neuron activity determines follower cell response
properties. \emph{J. Neurosci.} \textbf{32}, 2900 (2012). 

\bibitem {SMULKO2001}J.~Smulko, C-G. Granqvist, L.B. Kish. On the statistical
analysis of noise in chemical sensors and its application for sensing.
\emph{Fluct. Noise Lett.} \textbf{01}, L147 (2001). 

\bibitem {Scandurra2020}G.~Scandurra, J.~Smulko, L.B. Kish.
Fluctuation-enhanced sensing (FES): A promising sensing technique. \emph{Appl.
Sci.} \textbf{10}, 5818 (2020).\vspace*{-2mm} 
\begin{flushright}
{\footnotesize Translated from Ukrainian by O.I.~Voitenko}
\end{flushright}
\end{thebibliography}
\end{document}